# Surrogate Models to Predict Wave Hydrodynamics on Evolving Landscapes

Mohammad Ahmadi Gharehtoragh[1], David R Johnson[1,2]
[1] Edwardson School of Industrial Engineering, Purdue University, West Lafayette, IN
[2] Department of Political Science, Purdue University, West Lafayette, IN

## Abstract

Coastal planners using probabilistic risk assessments to evaluate structural flood risk reduction projects may wish to simulate the hydrodynamics associated with large suites of tropical cyclones in large ensembles of landscapes: with and without projects' implementation; over decades of their useful lifetimes; and under multiple scenarios reflecting uncertainty about sea level rise, land subsidence, and other factors. Wave action can be a substantial contributor to flood losses and overtopping of structural features like levees and floodwalls, but numerical methods solving for wave dynamics are computationally expensive, potentially limiting budget-constrained planning efforts. In this study, we present and evaluate the performance of deep learning-based surrogate models for predicting peak significant wave heights under a variety of relevant use cases: predicting waves with or without modeled peak storm surge as a feature, predicting wave heights while simultaneously predicting peak storm surge, or using storm surge predicted by another surrogate model as an input feature. All models incorporate landscape morphological elements (e.g., elevation, roughness, canopy) and global boundary conditions (e.g., sea level) in addition to tropical cyclone characteristics as predictive features to improve accuracy as landscapes evolve over time. Using simulations from Louisiana's 2023 Coastal Master Plan as a case study, we demonstrate suitable accuracy of surrogate models for planning-level studies, with a two-sided Kolmogorov-Smirnov test indicating no significant difference between significant wave heights generated by the Simulating Waves Nearshore model and those predicted by our surrogate models in approximately 89% of grid cells and landscapes evaluated in the study, with performance varying by landscape and model. On average, the models produced a root mean squared error of 0.05-0.06 m.

## Introduction

Storm surge, the abnormal rise of water during extreme storms above normal tides, is widely recognized as a highly hazardous event that can pose considerable risk to coastal areas and communities. Accurate prediction of storm surge, both from individual events and in a probabilistic sense, is crucial to inform risk management decisions and emergency response. However, risk reduction projects like levees and seawalls have multi-decadal useful lifetimes, meaning that planners need to evaluate how risk will evolve over time, introducing uncertainty about climate change, land subsidence, and other relevant factors. Further, advancing surge risk estimation demands a deeper understanding of physical processes, such as tidal effects and interactions between waves and storm surges, which can significantly strengthen flooding impacts in low-lying coastal zones (Staneva et al., 2016).

The significant wave height is defined as the average height of the highest third of the waves in a given period, or largest 33% of waves. Estimates of significant wave height exceedance probabilities are commonly used to inform planning, design and maintenance of coastal and offshore structures (American Society of Civil Engineers, 2022; Berbić et al., 2017; Mahjoobi & Adeli Mosabbeb, 2009; U.S. Army Corps of Engineers, 2008). Especially during extreme events like tropical cyclones, waves can cause several hazards such as coastal flooding and erosion that can lead to human loss and significant financial damages (Moghim et al., 2023). The impact of waves on storm surges can be significant especially during an extreme event; for example, Huang et al (2010) examined a coupled surge and waves model, finding that waves can incrementally increase the risk associated with storm surges and expand the footprint of coastal inundation. Near structural risk reduction systems like levees and floodwalls, wave action can be a major contributor to overtopping volumes and produce backside scour, leading to catastrophic failures. Integrating wave predictions with tidal and storm surge estimations provides a comprehensive approach to reducing flooding risks in coastal areas. (Merrifield et al., 2021; Phillips et al., 2017; Scott et al., 2020).

Waves can cause erosion in beaches and coastal areas, negatively affecting ecosystems and infrastructure (Harley et al., 2017; Huang et al., 2010; Narayan et al., 2016). Climate change also impacts landscape characteristics, such as reduced vegetation cover, loss of elevation, and reduction in horizontal extent, reduce the landscape's ability to mitigate the storm surge and wave impacts, also increase erosion, and alter bed roughness, which can lead to increased flooding (Wamsley et al., 2009). Further, Y. Yang et al (2015) demonstrated that another factor that has significant role in waves height is vegetation cover, that has notable effects on wave attenuation mechanisms and leads to a significant decrease in wave height. Similarly, two studies showed that vegetated foreshores, mangrove forests, and seagrass beds are capable of reducing wave loads and heights . Thus, there exist feedbacks between changes to landscape morphology and surge and wave hydrodynamics; Gharehtoragh & Johnson (2024) showed that these morphological parameters can be exploited to improve the prediction of peak storm surge as landscapes evolve and sea levels rise.

In recent decades, high-fidelity numerical models (i.e., physical-based models) have been developed to model storm surge and waves generated by hurricanes and tropical cyclones (TCs). However, in terms of computational cost, these models can be expensive and require substantial computing resources (Bilskie et al., 2014). Probabilistic flood risk assessments can demand simulation of a wide range of TC events with varying characteristics. Techniques, like the joint probability method with optimal sampling (JPM-OS), exist to reduce the number of required simulations by choosing a smaller and still representative set of TCs (Fischbach et al., 2016; Resio, 2007; Resio et al., 2009; Toro et al., 2010, 2010; K. Yang et al., 2019; J. Zhang et al., 2018). However, they still commonly prescribe running hundreds of TC events, which may be impossible for integrated planning studies considering future uncertainties.

Protection systems such as levees or flood walls, etc., need to resist extreme events over many decades, so designs should consider this uncertainty about future conditions. In the context of coastal flooding, uncertain parameters may include features of changing landscape morphology (e.g., land subsidence, land-use change, impacts of saltwater intrusion on vegetation) and boundary conditions (e.g., sea level rise). One way to address this issue is by employing scenario analysis that involves investigating future states of the world with different realizations of uncertain parameters (Kirwan et al., 2010; Sutton-Grier et al., 2018). Investigating multiple future states of the world to estimate future risk of extreme events requires a significant number of landscapes, and due to this, extensive computational resources are required, and employing more landscapes limits the number of events that can be simulated per landscape. On the other hand, using coarser resolutions in physically based models like ADCIRC (Advanced CIRCulation) or increasing mesh size could negatively affect the accuracy of the model.

One way to address this issue is by using surrogate models to predict storm surge and wave hydrodynamics (Kyprioti et al., 2021). In recent years, the utilization of surrogate models especially in water resource management field has increased (Asher et al., 2015; Razavi et al., 2012). Previous studies have used various ways to predict storm surge elevations and significant wave heights, primarily focusing on TC characteristics such as storm intensity and track parameters such as landfall location and heading, central pressure, forward velocity, radius of maximum wind speed, Holland-B parameter and/or tide level. For instance, Deo et al (2001) used an artificial neural network (ANN) to predict significant wave heights utilizing TC parameters such as wind speed and directions. Vijayan et al. (2023) employed the dynamically-coupled ADCIRC+SWAN model to predict waves, showing the relationship between sea level rise and wave heights. Similarly Londhe and Panchang (2006) utilized an ANN for one-day wave forecasting, showing that they can be useful for wave prediction but may be less accurate in predicting the magnitudes of the highest waves. Some recent studies such as Zhang et al. (2021) employed numerical long short-term memory frameworks to predict wave heights using a combination of a current wave measurement and a numerical prediction from the Simulating Waves Nearshore (SWAN) model.

In this paper, we introduce knowledge-guided surrogate models utilizing artificial neural networks (ANNs) that are able to alleviate the computational burdens of predicting peak significant wave heights. This model was trained based on synthetic tropical cyclone (TC) data produced by a coupled ADCIRC+SWAN (Simulating WAves Nearshore) model on multiple landscapes representing projected conditions in coastal Louisiana from a 2020 baseline over decadal time slices through 2070. In addition to predicting wave heights as a function of TC parameters at landfall, we also employ morphological features (e.g., topographic/bathymetric elevations, roughness) and global boundary conditions (e.g., mean sea level), allowing us to train models on simulations from multiple landscapes simultaneously.

Moreover, we evaluate four models designed to represent different potential real-world use cases. Recent versions of ADCIRC are much faster at solving storm surge hydrodynamics, but the gains are only realized when not coupled to a wave model. Thus, a baseline model utilizing only TC landfall, landscape, and sea level data as features is compared to another model which also includes ADCIRC-simulated peak storm surge as a feature. This second model could be used in applications where ADCIRC is run in uncoupled mode but generating waves is still desirable. Of course, policy makers may still wish to run experimental designs that exceed their computational budget to support adaptive planning efforts or methods for decision making under deep uncertainty. In this case, similar surrogate models may be useful for predicting peak storm surge as well, so we evaluate a model of peak significant wave heights that includes a surrogate model-predicted peak storm surge as a feature. Finally, we compare this to a fourth model that is trained to predict peak storm surge and significant wave heights simultaneously. In addition to evaluating predictions associated with individual TCs, we statistically aggregate the predictions for TCs on each landscape to produce estimates of annual exceedance probability (AEP) distributions.

## Methods

### *Data Description*

Synthetic tropical cyclones (i.e., TCs following an idealized, regular track and patterns of intensification and decay) are used in this study, with each synthetic storm parameterized by their forward velocity, radius of maximum wind speed, central pressure, landfall coordinates, and heading. The corpus of 645 synthetic TCs used in this analysis was created using the JPM-OS methodology for flood risk assessments in Louisiana (Nadal-Caraballo et al., 2020, 2022). Table S1 provides a complete list of the landfall parameters for each synthetic storm. These parameters serve as input data for all of the predictive models.

For all 645 synthetic storms, hydrodynamic simulations were provided by a coupled ADCIRC+SWAN model from Louisiana's 2023 Coastal Master Plan that was simulated on the plan's "Existing Conditions" landscape (i.e., 2020) (Louisiana Coastal Protection and Restoration Authority, 2023b). Uncertainty in future conditions is represented in the plan by "Lower" and "Higher" environmental scenarios which vary in their assumptions about sea level rise, land subsidence, changes to TC intensity, and other environmental factors (Cobell & Roberts, 2021; Louisiana Coastal Protection and Restoration Authority, 2023a). A subset of 90 synthetic storms were simulated on each of ten future landscapes representing decadal snapshots of the Lower and Higher scenarios from 2030 to 2070; these 90 storms are bolded in Table S1 for reference. Peak significant wave heights were extracted from the coupled ADCIRC+SWAN outputs at 80,224 locations that comprise the Coastal Louisiana Risk Assessment (CLARA) model's grid points in Louisiana that are not located inside fully-enclosed risk reduction systems (i.e., ring levees and floodwalls). These points form a mixed-resolution mesh with a maximum 1-km spacing and higher resolution in some areas, such that every U.S. Census block contains at least one grid point (Johnson et al., 2023).

Each landscape is characterized spatially by digital elevation models (DEMs) of topography and bathymetry, and average slope of nearby cells at each location, and by GeoTIFF rasters of free surface roughness $z0$, Manning's $n$ values (i.e., bottom roughness coefficient), and a surface canopy coefficient that captures the reduction in wind stress on water surfaces produced by local vegetation. All landscape characteristics were extracted for all 80,224 locations in the study area to be used in the developed surrogate model. All required information about the Integrated Compartment Model employed to develop the landscape representations can be found in White et al. (2019) and Reed and White (2023), and details regarding the ADCIRC+SWAN model and Louisiana mesh are found in Cobell and Roberts (2021) and Roberts and Cobell (2017).

## *Model Development*

In this study, four different models were developed to predict peak significant wave height. All models employed the same landscape parameters, including latitudinal and longitudinal coordinates, topo/bathymetry elevation, surface canopy, Manning's $n$ coefficient, $z0$, and average slope (estimated by calculating the average slope of the adjacent cells using DEM values). TC landfall parameters used as features included forward velocity, radius of maximum wind speed, central pressure, landfall coordinates, and heading. Finally, a global boundary condition of mean sea level in each landscape was also used as a feature.

These four models differ in their final input features and targets. The first model (Model 1), used as a baseline, predicts waves independently of surge, i.e., using only the TC landfall parameters, landscape features, and mean sea level. Model 2 includes as a feature the peak storm surge predicted by a surge surrogate model. Model 3 instead utilizes peak storm surge elevations simulated by the ADCIRC+SWAN model. Lastly, Model 4 predicts both peak storm surge elevations and significant wave heights simultaneously.

The surrogate model configuration consists of a Convolutional Neural Network (CNN) followed by multiple dense layers. The developed CNN models consisted of several convolutional layers, each one of them containing a range of 128 filters to 256 filters, followed by batch normalization layers, dropout layers, and RELU activation functions. Further, to address vanishing gradient and allow information to flow across layers and train a deeper model, three skip connections were implemented to pass the information from early layers to the last layers.

In addition, four dense layers that include a range of 128 to 256 neurons were defined. In defining the number of neurons, one should be cautious, since having too few neurons and filters could prevent models from training correctly, and having too many of them could result in overtraining/overfitting (Albawi et al., 2017; Alemu et al., 2018). Moreover, a similar skip connection is also implemented in the dense layers too, to pass the information from the first layer directly to all next layers, with the same for subsequent layers.

Further, for all layers including convolution and dense layers, the RELU activation function was selected with the callback approach that has an adaptive learning rate starting at 0.01; if validation loss does not decrease in two time steps, it reduces learning rate by a factor of 0.75, potentially going to a minimum value of 0.00001 (Smith, 2017). In the last layer, a linear activation function was implemented to predict wave values at each location. For all models except Model 4, an output layer of 1 dimension was used. However, for Model 4 a 2-dimensional output layer was utilized to predict both peak wave and surge height simultaneously. The entire simulation was executed on GPU resources (Nvidia A100) taking less than 5 hours for training the models; inference on a new landscape takes less than 5 minutes.

To have a more realistic evaluation of model accuracy, leave-one-out cross-validation (LOOCV) was performed on the future landscapes. In other words, each time a model was trained, 1 of the 10 future landscapes ($n = 90$ storms) was dropped and used as a test set, and the rest of the 9 future landscapes along with the 2020 landscape ($n = 1455$ storms) were used as a training set. The 2020 landscape was consistently utilized in training throughout the entire process. This procedure represents a potential use case of the surrogate model as a scenario generator to produce predictions for many TCs on a novel landscape, as opposed to alternative cross-validation procedures that would drop a fraction of storms or grid points from multiple landscapes.

Lastly, planners and project designers need to know how errors in predicting peak wave height manifest as differences in the estimated annual exceedance probability (AEP) distribution for wave heights in each landscape. This study leveraged the CLARA model to calculate wave hazard curves (i.e., AEP distributions) associated with the surrogate model predictions for each synthetic TC (Johnson et al., 2013, 2023). The complete methodology of the CLARA model and further details can be found in Johnson et al. (2023). The model was used to calculate wave height exceedances at 23 different return periods, covering from 50% AEP to 0.005% AEP. Further, by utilizing the CLARA model, exceedance curves (wave heights as a function of AEP) were generated from both simulated peak significant wave height of the surrogate model through the LOOCV procedure and the simulated peak significant wave height from ADCIRC model. As a final step, the empirical distributions generated from hazard curves using ADCIRC and surrogate models were compared using a two-sided Kolmogorov–Smirnov (K-S) test. The K-S test measures the maximum absolute difference between the two empirical cumulative distributions. The null hypothesis is that the two samples are drawn from the same underlying distribution. So, the K-S test can be applied at each grid point to determine how many points reject the null hypothesis at significance level of $\alpha = 0.05$.

$$\sup_{x}|F_{ADCIRC}(x) - F_{ANN}(x)| > \sqrt{-\ln\left(\frac{\alpha}{2}\right) \cdot \frac{1}{23}} \tag{1}$$

In the above formula, sup is the supremum over x, $F_{ANN}(x)$ is the sample CDF associated with the surrogate model predictions, and $F_{ADCIRC}(x)$ the CDF associated with the ADCIRC simulations.

## Results

As expected intuitively, the use of simulated surge elevations from ADCIRC as an additional training feature led Model 3 to generally out-perform the other models with respect to root mean squared error (RMSE) over the grid points, landscapes, and synthetic storms (Figure 1). Each point in the scatter plot indicates the RMSE of an individual grid point over all landscapes and synthetic storms for a given model (e.g., Model 3 colored in red) on the horizontal axis, with the vertical placement indicating the percentage of grid points with RMSE equal to or exceeding the given value; as an example of how this is interpreted, the figure indicates that for Model 3, only 0.1% percent of grid cells have an RMSE equal to or exceeding approximately 0.32 m. Thus, plots farther toward the top-left of the figure indicate better overall accuracy.

The performance of the baseline Model 1 and Model 4 that predicts both peak surge height and significant wave height is approximately the same across all ranges of RMSE values. This was expected since Model 1, the baseline model, was trained and focused on predicting only significant wave height. Model 4 predicts both significant wave height and peak surge elevation, benefiting from the informative relationship between these two variables. Lastly, Model 2, utilizing surge elevations predicted by a surrogate model rather than ADCIRC, shows slightly weaker performance among all models and this difference can be seen particularly for RMSE values greater than 0.2 meters. That is likely associated with compounding errors and biases from the surge surrogate model, on top of biases or noise in the underlying ADCIRC simulations.

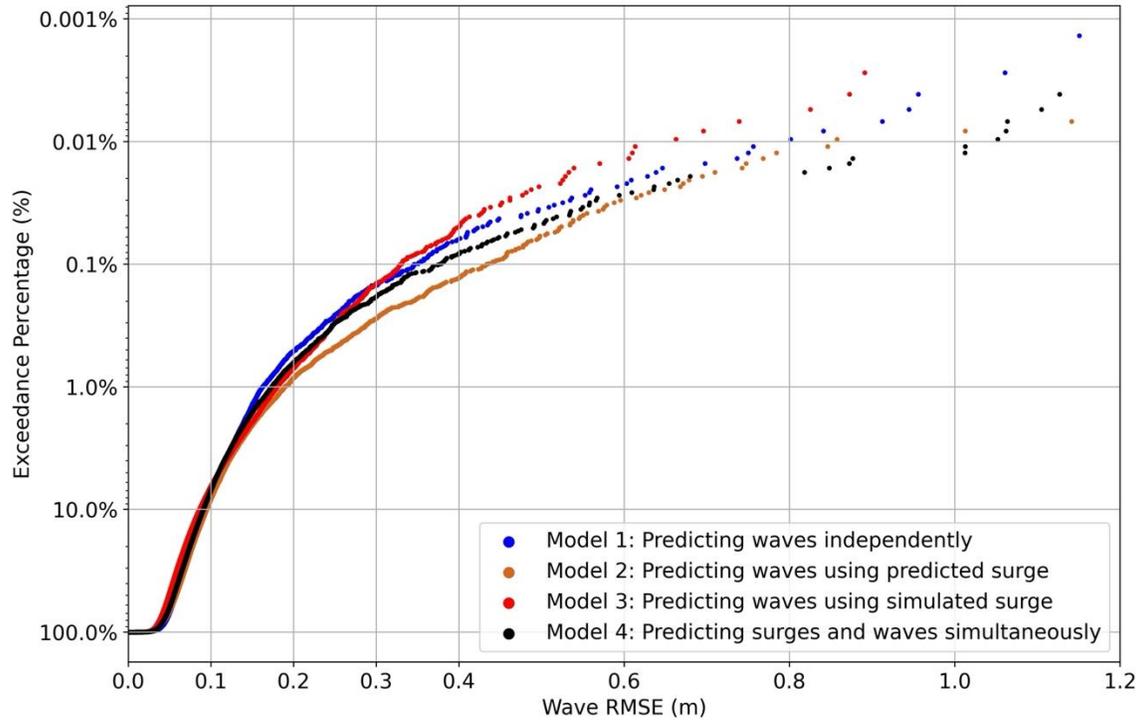

Figure 1. Exceedance percentage of RMSE values by grid point and model, with RMSE averaged across all landscapes and synthetic storms.

For the baseline Model 1, the RMSE at 90% of grid points is less than 0.09 m, at 99% of grid points less than 0.16 m, and at 99.9% of grid points less than 0.34 m (Figure 1). Including the simulated ADCIRC surge as a feature in Model 3 provides additional explanatory accuracy relative to the baseline, so due to that Model 3 showed best performance at the 99.9% of the grid points with RMSE value of less than 0.32 m, while the predicted surge used by Model 2 apparently compounds errors in the predicted wave dynamics instead of being accurate enough to yield improvements compared to Model 1. Similar to the baseline Model 1, Model 4 does not use surge data as an input, however, compared to baseline Model 1, Model 4 predicts both peak surge height and significant wave height, so as was expected, it closely mirrored the performance of Model 1 in its prediction of the wave height outcomes. By and large, besides Model 2, which uses predicted surge data and showed weaker performance than other models, the remaining models showed similar performance and were sufficiently accurate for use in planning-level studies over a large coastal domain (i.e., relatively few outliers with large RMSE compared to the RMSE of the underlying ADCIRC model).

Model 3, by using simulated surge data, was capable of performing better in comparison to other models and showed an average RMSE of around 0.04-0.05 m over all landscapes except the Higher scenario in 2070 (Table *1*). Additionally, Model 4, despite predicting both surge and wave simultaneously without using surge as a feature, was able to predict surge elevations with RMSE less than 0.1 m in all landscapes but the 2070 Higher scenario. Across all four models, we see substantially degraded accuracy in the Higher scenario's 2070 landscape. This is a byproduct of the 2023 Coastal Master Plan's experimental design, where sea level rise is assumed to accelerate over time, leading the 2070 Higher landscape to be the "most different" case relative to any other landscape in the training data, both with respect to the sea level boundary condition and other morphological features. Consequently, predicting hydrodynamics is a fundamentally more challenging extrapolation problem in the 2070 Higher scenario landscape than in others.

The RMSE values of the baseline Model 1, and differences in RMSE for other models compared to the baseline are shown in Figure 2 (grid cells with absolute differences less than 0.025 m not shown). Warmer yellow-to-red colors indicate a model's improvement relative to the baseline model, while cool green-to-violet colors indicate worse accuracy at that grid point.

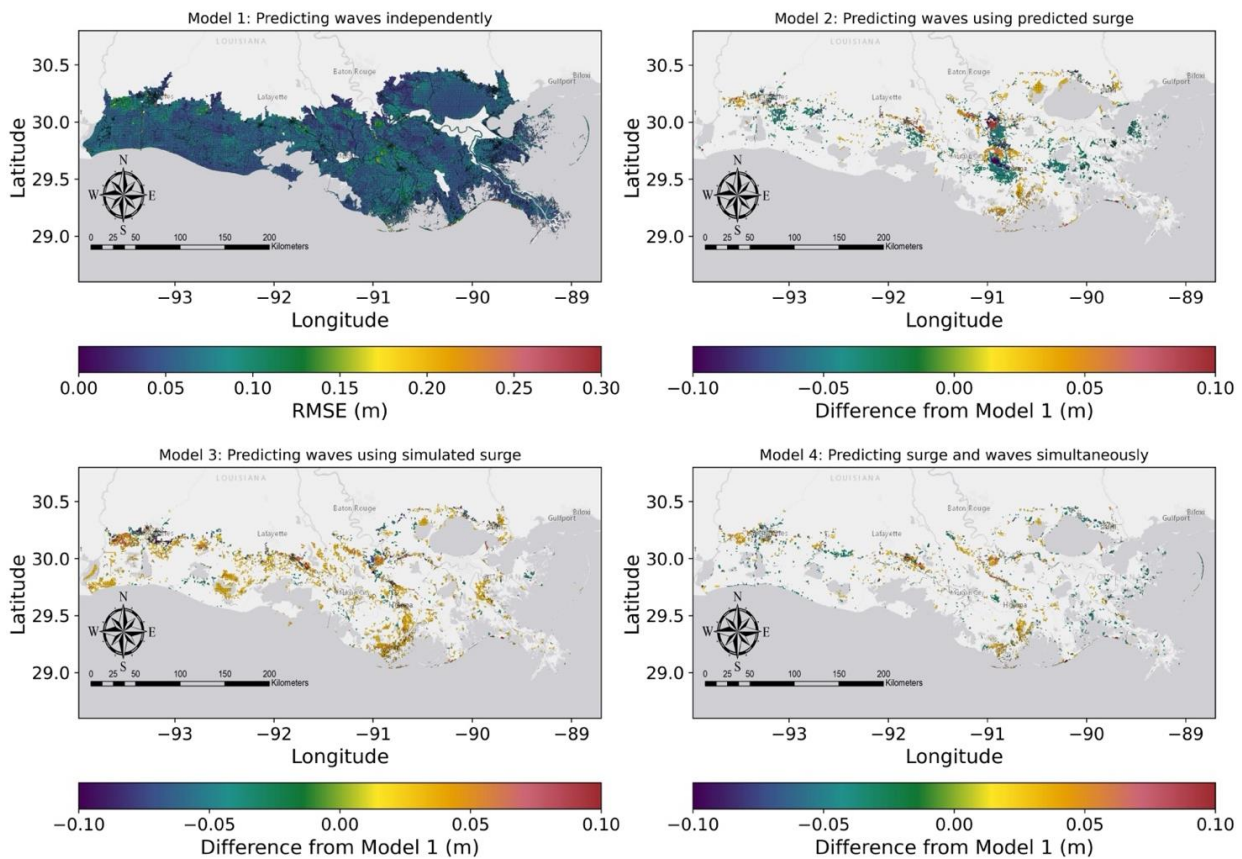

Figure 2. RMSE of Model 1 (top left pane) and difference maps for the other models relative to Model 1 (i.e., Model 1 value minus Model 2/3/4 value) across all landscapes and synthetic storms for all grid points in the study domain (differences within ±0.025 m excluded for contrast).

Consequently, we see that the general improvements in RMSE for Model 3 are widespread across the coastal zone, with limited areas with worse accuracy near the inland extent of the model domain, such as in Lake Charles in the northwestern part of the coastal zone. Model 2 has a similar number of grid cells where differences exceed ±0.025 m, but those differences are generally worse performance than the baseline Model 1, especially in the wetlands of the Atchafalaya River basin. Model 4 shows the greatest similarity to the baseline, with considerably fewer grid cells exhibiting differences beyond ±0.025 m. In Table *1*, we show a summary of the statistical metrics (RMSE and Pearson correlation coefficient) used to evaluate the models.

Table *1* Summary of statistical outcomes (RMSE and correlation coefficient) for all cases evaluated.

| Scenarios | Years | Model 1 | | Model 2 | | Model 3 | | Model 4 | | | |
|---|---|---|---|---|---|---|---|---|---|---|---|
| | | Wave | | Wave | | Wave | | Surge | | Wave | |
| | | RMSE (m) | Corr | RMSE (m) | Corr | RMSE (m) | Corr | RMSE (m) | Corr | RMSE (m) | Corr |
| Lower Scenario | 2030 | 0.048 | 0.995 | 0.048 | 0.995 | 0.042 | 0.996 | 0.063 | 0.998 | 0.055 | 0.993 |
| | 2040 | 0.047 | 0.995 | 0.050 | 0.995 | 0.043 | 0.996 | 0.093 | 0.996 | 0.054 | 0.994 |
| | 2050 | 0.049 | 0.995 | 0.053 | 0.994 | 0.044 | 0.996 | 0.084 | 0.997 | 0.054 | 0.994 |
| | 2060 | 0.050 | 0.995 | 0.054 | 0.994 | 0.046 | 0.996 | 0.072 | 0.998 | 0.057 | 0.993 |
| | 2070 | 0.064 | 0.993 | 0.066 | 0.992 | 0.054 | 0.995 | 0.088 | 0.996 | 0.064 | 0.992 |
| Higher Scenario | 2030 | 0.044 | 0.995 | 0.045 | 0.995 | 0.042 | 0.996 | 0.055 | 0.999 | 0.052 | 0.994 |
| | 2040 | 0.047 | 0.995 | 0.048 | 0.995 | 0.043 | 0.996 | 0.072 | 0.997 | 0.054 | 0.994 |
| | 2050 | 0.050 | 0.995 | 0.066 | 0.994 | 0.050 | 0.995 | 0.086 | 0.997 | 0.062 | 0.993 |
| | 2060 | 0.057 | 0.994 | 0.060 | 0.993 | 0.051 | 0.995 | 0.082 | 0.997 | 0.062 | 0.993 |
| | 2070 | 0.121 | 0.980 | 0.112 | 0.980 | 0.103 | 0.982 | 0.172 | 0.988 | 0.093 | 0.985 |

Based on the results of Table *1* and maps that reveal no troubling spatial patterns of bias, we view all four models as being accurate enough for use in planning studies. Each has a potential use case that would depend on the computational resources available for use in generating training data using more expensive models like ADCIRC+SWAN.

This conclusion is bolstered by aggregating the wave heights generated by individual synthetic TCs to estimate a hazard curve. Figure 3 shows RMSE values across all grid points, grouped by model in each column, landscape (Lower and Higher scenarios on top and bottom rows, respectively, with the year in different colors), and annual exceedance probability (AEP). The horizontal axes within each pane shows the AEP, and the vertical axis locates the RMSE values.

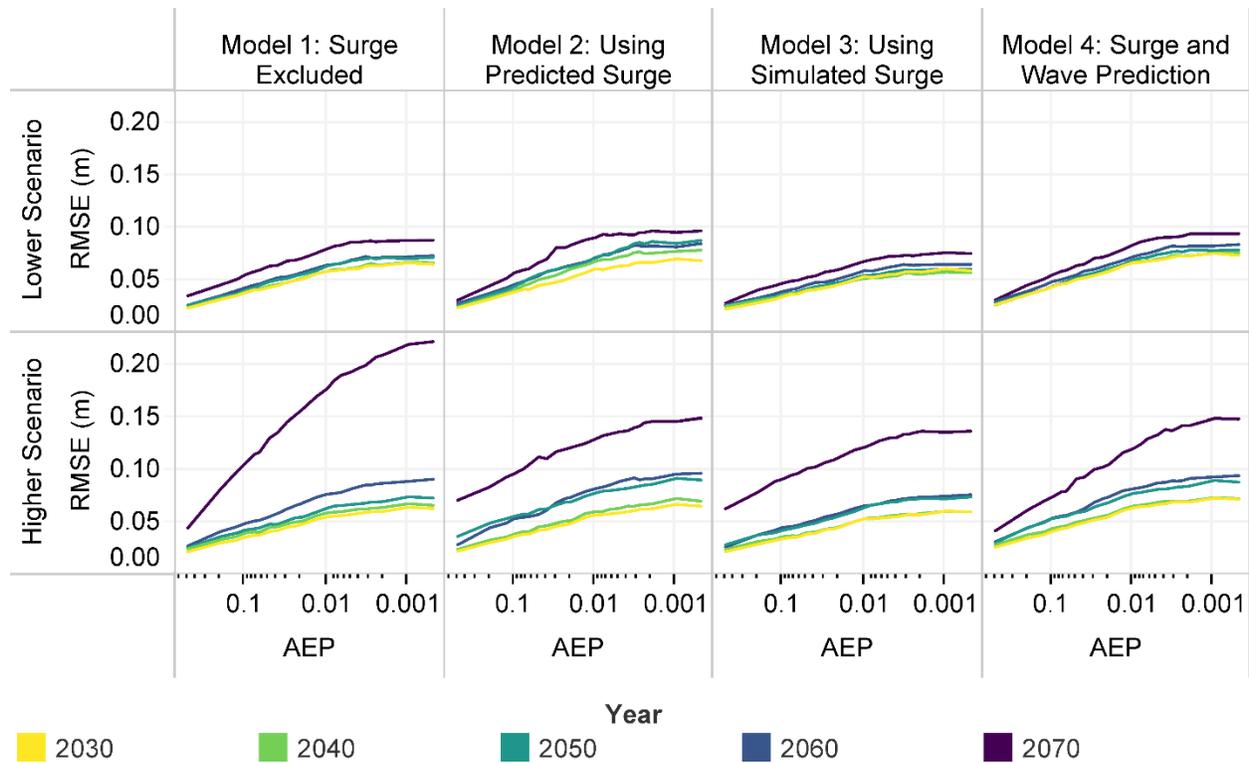

Figure 3. RMSE over all grid points, by model, annual exceedance probability, and landscape.

The results suggest that the surrogate models have broadly similar accuracy of 0.02-0.1 m across the range of synthetic storms that produce peak significant wave heights associated with a wide range of return periods, from the 0.5 AEP (i.e., 2-year) to 0.0005 AEP (i.e., 2,000-year) events. It is, however, more challenging to predict extremes, consistent with higher RMSE values farther into the tail of the distribution. To ensure this would not pose an issue for studies focused on extreme events, i.e., 100-year or 0.01% AEP and beyond, we also examine the normalized RMSE (NRMSE) in Figure 4. It shows that the average error is generally less than 2% in each of the various landscapes and across all return periods, with the exception of the 2070 Higher Scenario landscape. This landscape continued to pose a challenge to all models due to its inherent extrapolation when implementing the leave-one-landscape-out cross-validation procedure.

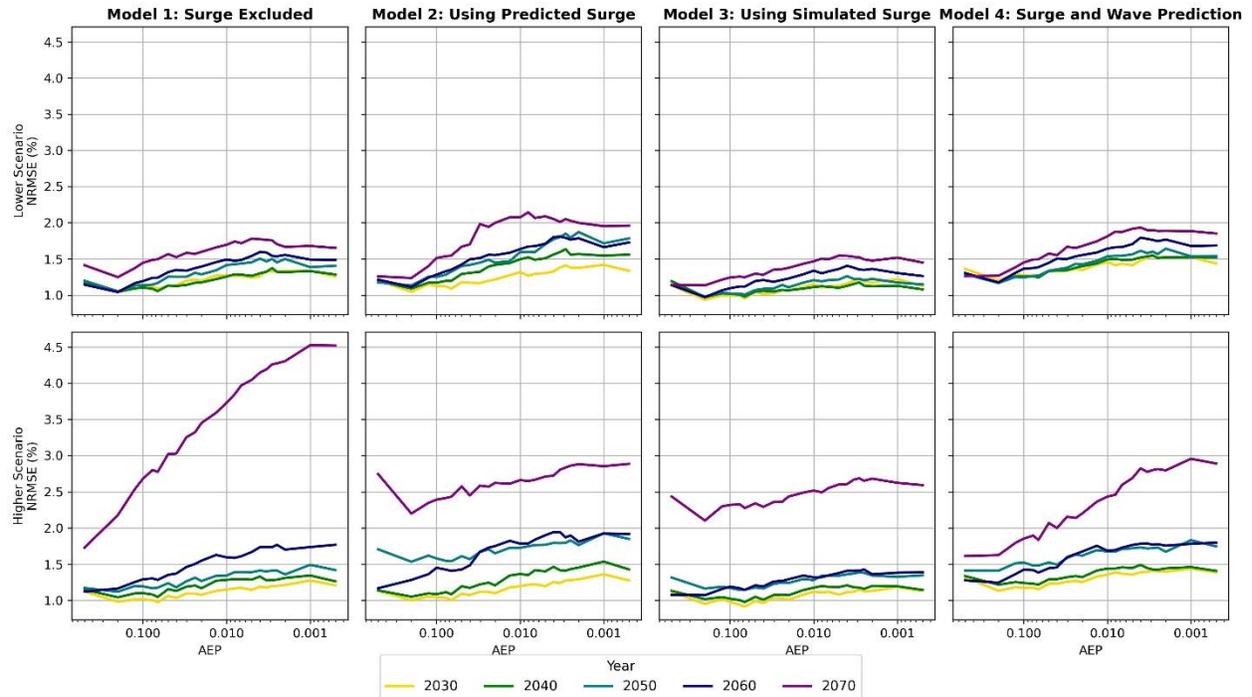

Figure 4. Normalized RMSE over all grid points, by model, annual exceedance probability, and landscape.

The small NMRSE values by return period when averaging over all grid points suggested that the surrogate models do a good job of emulating the hazard curves produced by applying the JPM-OS methodology to the original ADCIRC+SWAN simulations. Prior studies show the potential for accuracy in emulating surge hazard curves, but here we see some differences in performance across the four models producing wave height estimates. The 2070 Higher Scenario landscape again is an outlier in average performance on the NRMSE metric, but notably worse so for the baseline Model 1; Model 4, predicting surge and waves simultaneously, also performed notably better on this landscape for more frequent return periods, i.e., higher AEP exceedances. Expanding beyond the average results, Table 2 shows the comparison between the AEP distributions calculated from all four models and the AEP distributions obtained from ADCIRC predictions using the two-sample K–S test with a significance level of $\alpha = 0.05$. Values in the table indicate the percentage of grid points where the null hypothesis, that the surrogate model ADCIRC hazard curves are drawn from the same distribution over the 23 return periods estimated. Points where the null hypothesis is rejected represent outliers where the wave height predictions result in statistically significant differences in estimated hazard distributions.

Table 2 Percentage of grid points rejecting a two-sided KS test null hypothesis over each model per landscape

| Scenario | Year | Model 1 Rejected % | Model 2 Rejected % | Model 3 Rejected % | Model 4 Rejected % |
|---|---|---|---|---|---|
| Lower Scenario | 2030 | 7% | 7% | 5% | 8% |
|  | 2040 | 7% | 9% | 6% | 9% |
|  | 2050 | 7% | 9% | 6% | 9% |
|  | 2060 | 8% | 9% | 7% | 10% |
|  | 2070 | 11% | 13% | 10% | 13% |
| Higher Scenario | 2030 | 6% | 7% | 6% | 8% |
|  | 2040 | 8% | 9% | 6% | 9% |
|  | 2050 | 9% | 13% | 8% | 12% |
|  | 2060 | 11% | 13% | 8% | 12% |
|  | 2070 | 48% | 24% | 22% | 24% |

Model 3, by using simulated surge data, achieved the smallest percentage of rejected points over all 10 future landscapes. Models 2 and 4 show similar performance with slightly higher percentages compared to Model 3. The baseline Model 1 performance in the 2070 Higher Scenario landscape is by far the worst statistical outcome, with nearly half of points failing to successfully emulate the ADCIRC hazard distribution with $\alpha = 0.05$, though all models showed their worst performance by far on this landscape.

## Discussion

In this study four machine learning-based surrogate models were developed to predict peak significant wave heights. The developed models are capable of predicting either wave or surge/wave dynamics simultaneously with accuracy comparable to that of the calibration and validation accuracy of the underlying ADCIRC+SWAN model simulations.

These models demonstrated that by utilizing future landscapes with varying landscape parameters, average slope and mean sea level conditions, their accuracy could be improved relative to current models that focus on predicting hydrodynamics on static landscapes. By utilizing LOOCV in this study, one future scenario was left out each time as test data, and the models were able to predict wave heights with an approximate RMSE of 0.05–0.06 m, with Model 2 generally demonstrating slightly worse performance across the range of metrics evaluated.

Further, we utilized a two-sided K-S test on the hazard curves generated by ADCIRC+SWAN simulations and the surrogate models at each grid point to assess whether the samples were drawn from statistically different underlying distributions. The results of Table 1 and Table 2 showed that, for all cases, on average, less than 8% of grid points rejected the null hypothesis, except for the more extreme 2070 landscape. The best performance came from Model 3, showing how the use of surge elevations are informative to making more accurate predictions under these extreme conditions. Our results did not show bias toward underestimation or overestimation.

We have mentioned several times that, by all measures, the models are less accurate in the Higher Scenario's 2070 landscape, the most extreme case with highest sea level rise. This illustrates a limitation of using training data of convenience, i.e., simulations that were already available for methods development from Louisiana's 2023 Coastal Master Plan. An important implication is that when surrogate models are being considered for deployment in a planning study, this should inform the experimental design for what should be simulated and used as training data. For example, the Coastal Master Plan uses a 50-year planning horizon, so decisions are made based in part on risk estimates in 2070; surrogate model accuracy for the Higher Scenario's 2070 landscape would likely improve if ADCIRC+SWAN simulations were available for either a 2080 landscape or a 2070 landscape under a scenario with still-higher sea level rise. As a reminder, the same 90 storms were simulated in each future landscape based on a storm selection process that considered only the baseline 2020 landscape (Fischbach et al., 2021). Future research could examine the use of adaptive sampling techniques for surrogate model training in this specific context. Instead of simulating the same 90 storms as were performed for the Coastal Master Plan, sampling different synthetic storm events for each landscape could be more effective in terms of computational efficiency.

Another key point is that surrogate models are capable of predicting both surge and wave dynamics simultaneously with sufficient accuracy. The possible logic behind this is the physical relationship between surge elevations and significant wave heights that helps the model to make more realistic predictions of both parameters. While the model architectures employed in this analysis were not physics-constrained, the models are still potentially able to learn the impact on wave heights of breaking behaviors induced by surge depth limitations or sloping topography. Model 4 achieves acceptable performance without adding a considerable amount of computational cost, making it suitable as a scenario generator for a wide range of possible futures.

The training data of this study came from the Coastal Master Plan's Future Without Action scenarios, meaning that these scenarios are limited to only slowly evolving landscapes, and no further projects—like upgrading levee systems or floodwalls—have been implemented. Although developing a machine learning framework to consider these types of linear weir features is more challenging, coastal restoration projects that affect the morphology of future landscapes can still be examined using these surrogate models. By having low computational cost, they can be used as a scenario generator, and thousands of future landscape scenarios could be evaluated, including those that reflect coastal restoration projects like marsh creation, river diversions, and barrier island replenishment. Each of the four models presented here have potential use cases for planning studies with varying computational budgets and decision frameworks, enabling the use of methods for optimization or decision-making under deep uncertainty that require a large ensemble of future states of the world, a large number of function evaluations, or both.

## Acknowledgements

This work was supported by the U.S. National Science Foundation under awards 2238060 and 2118329. We thank members of the modeling teams for Louisiana's Comprehensive Master Plan for a Sustainable Coast and funding from the Coastal Protection and Restoration Authority, under the 2023 and 2029 Coastal Master Plan's Master Services Agreement, for the original simulation data used in this study. Any opinions, findings, conclusions, and recommendations expressed in this material are those of the authors and do not necessarily reflect the views of the funding entities.

# Supplementary Information

Table S1. Distribution of synthetic storm parameters at landfall.

| Storm ID | Heading | $v_f$ (knots) | $r_{max}$ (nm) | Landfall lon ($x$) | $c_p$ (mbar) |
|---|---|---|---|---|---|
| 1 | 35.8 | 9.5 | 10.9 | -102.376 | 865.25 |
| 2 | 35.8 | 15.3 | 14.7 | -102.375 | 885.25 |
| 3 | 35.8 | 11.8 | 5.9 | -102.377 | 905.25 |
| 4 | 35.8 | 9.1 | 9.2 | -102.377 | 925.25 |
| 5 | 35.8 | 16.7 | 15.5 | -102.378 | 945.25 |
| 6 | 35.8 | 8.3 | 31.3 | -102.376 | 965.25 |
| 7 | 35.8 | 10.2 | 59 | -102.377 | 985.25 |
| 8 | 35.8 | 9.9 | 23.4 | -102.377 | 1005.25 |
| 9 | 62.72727 | 20.6 | 9 | -98.8967 | 865.25 |
| 10 | 62.72727 | 7.3 | 5.1 | -98.8991 | 885.25 |
| 11 | 62.72727 | 8.6 | 27.3 | -98.8964 | 905.25 |
| 12 | 62.72727 | 10.2 | 25.3 | -98.8974 | 925.25 |
| 13 | 62.72727 | 4.8 | 27.5 | -98.8975 | 945.25 |
| 14 | 62.72727 | 9.3 | 22.4 | -98.8986 | 965.25 |
| 15 | 62.72727 | 9.7 | 11.7 | -98.899 | 985.25 |
| 16 | 62.72727 | 10.9 | 44.5 | -98.899 | 1005.25 |
| 17 | 69.86364 | 9.8 | 5 | -95.3612 | 865.25 |
| 18 | 69.86364 | 14.4 | 11.9 | -95.3631 | 885.25 |
| 19 | 69.86364 | 5.1 | 16.4 | -95.36 | 905.25 |
| 20 | 69.86364 | 17.2 | 10.2 | -95.359 | 925.25 |
| 21 | 69.86364 | 7.8 | 36.8 | -95.3605 | 945.25 |
| 22 | 69.86364 | 9.8 | 25.1 | -95.3612 | 965.25 |
| 23 | 69.86364 | 4.6 | 9 | -95.3626 | 985.25 |
| 24 | 69.86364 | 12.3 | 56.6 | -95.3634 | 1005.25 |
| 25 | 69.88333 | 10.9 | 6 | -91.8512 | 865.25 |
| 26 | 69.88333 | 5.2 | 8 | -91.85 | 885.25 |
| 27 | 69.88333 | 10.5 | 19.8 | -91.85 | 905.25 |
| 28 | 69.88333 | 6.7 | 42.3 | -91.8499 | 925.25 |
| 29 | 69.88333 | 17.5 | 26.5 | -91.8528 | 945.25 |
| 30 | 69.88333 | 8.6 | 11.4 | -91.8494 | 965.25 |
| 31 | 69.88333 | 11.8 | 51.2 | -91.8525 | 985.25 |
| 32 | 69.88333 | 5.2 | 35.9 | -91.8503 | 1005.25 |
| 33 | 77.34848 | 11.2 | 7.7 | -88.3561 | 865.25 |
| 34 | 77.34848 | 18.1 | 15.7 | -88.3545 | 885.25 |
| 35 | 77.34848 | 11.4 | 17.9 | -88.3557 | 905.25 |
| 36 | 77.34848 | 8.5 | 11.8 | -88.3535 | 925.25 |
| 37 | 77.34848 | 9.3 | 49.6 | -88.3521 | 945.25 |
| 38 | 77.34848 | 5 | 27.1 | -88.3541 | 965.25 |
| 39 | 77.34848 | 4.5 | 19 | -88.3534 | 985.25 |
| 40 | 77.34848 | 13.2 | 33.4 | -88.355 | 1005.25 |
| 41 | 88.96364 | 6.3 | 8.8 | -85.1718 | 865.25 |
| 42 | 89.01786 | 11.3 | 6.3 | -85.1751 | 885.25 |
| 43 | 88.96364 | 13.2 | 29 | -85.1714 | 905.25 |
| 44 | 88.96364 | 8 | 34.8 | -85.1727 | 925.25 |
| 45 | 88.96364 | 6.2 | 17.6 | -85.1724 | 945.25 |
| 46 | 88.96364 | 17 | 23.3 | -85.1735 | 965.25 |
| 47 | 88.96364 | 8.5 | 21.9 | -85.1722 | 985.25 |
| 48 | 88.96364 | 7.7 | 62.4 | -85.1726 | 1005.25 |
| 49 | 48.39024 | 9.6 | 17.7 | -96.13 | 875.25 |
| 50 | 48.39024 | 15.9 | 16.4 | -96.1307 | 895.25 |
| 51 | 48.39024 | 8.7 | 8.9 | -96.1314 | 915.25 |
| 52 | 48.39024 | 9.6 | 9.3 | -96.13 | 935.25 |
| 53 | 48.39024 | 16.8 | 13 | -96.1303 | 955.25 |
| 54 | 48.39024 | 11.5 | 50.9 | -96.131 | 975.25 |
| 55 | 48.39024 | 4.8 | 32.6 | -96.1309 | 995.25 |
| 56 | 49.06522 | 21.9 | 13.9 | -94.9388 | 875.25 |
| 57 | 49.06522 | 14.9 | 9.4 | -94.9391 | 895.25 |
| 58 | 49.06522 | 9.3 | 7.6 | -94.9392 | 915.25 |
| 59 | 49.06522 | 10.5 | 22.4 | -94.9383 | 935.25 |
| 60 | 49.06522 | 5.3 | 16 | -94.9385 | 955.25 |
| 61 | 49.06522 | 11.9 | 48.7 | -94.9389 | 975.25 |
| 62 | 49.06522 | 12 | 15 | -94.9388 | 995.25 |
| 63 | 51.03846 | 24.4 | 7.6 | -93.7258 | 875.25 |
| 64 | 51.03846 | 13.2 | 9.7 | -93.7262 | 895.25 |
| 65 | 51.03846 | 5.5 | 19.4 | -93.7253 | 915.25 |
| **66** | **51.03846** | **8** | **44.4** | **-93.7255** | **935.25** |
| 67 | 51.03846 | 8.3 | 9.2 | -93.7247 | 955.25 |
| **68** | **51.03846** | **12.6** | **35.9** | **-93.7253** | **975.25** |
| 69 | 51.03846 | 8.4 | 28 | -93.725 | 995.25 |
| 70 | 51.05172 | 5.3 | 13.4 | -92.5434 | 875.25 |
| 71 | 51.05172 | 17.5 | 11.8 | -92.5442 | 895.25 |
| 72 | 51.05172 | 9.6 | 9.7 | -92.5436 | 915.25 |
| 73 | 51.05172 | 4.9 | 15.4 | -92.5437 | 935.25 |
| 74 | 51.05172 | 5.8 | 27 | -92.5433 | 955.25 |
| 75 | 51.05172 | 9.8 | 44.8 | -92.5445 | 975.25 |
| 76 | 51.05172 | 6 | 11.2 | -92.5439 | 995.25 |
| 77 | 52.05085 | 13.6 | 11.2 | -91.324 | 875.25 |
| 78 | 52.05085 | 9.2 | 13.2 | -91.325 | 895.25 |
| 79 | 52.05085 | 14.5 | 20.7 | -91.3244 | 915.25 |
| **80** | **52.05085** | **9.3** | **48** | **-91.3253** | **935.25** |
| 81 | 52.05085 | 9.6 | 12.2 | -91.3248 | 955.25 |
| **82** | **52.05085** | **6.2** | **28.8** | **-91.324** | **975.25** |
| **83** | **52.05085** | **15** | **47** | **-91.3246** | **995.25** |
| 84 | 54.83051 | 6.8 | 13 | -90.1105 | 875.25 |
| 85 | 54.83051 | 15.4 | 12.9 | -90.1105 | 895.25 |
| 86 | 54.83051 | 12.5 | 14.4 | -90.1102 | 915.25 |
| **87** | **54.83051** | **9.1** | **41.7** | **-90.11** | **935.25** |
| 88 | 54.83051 | 13.6 | 22.5 | -90.1105 | 955.25 |
| 89 | 54.83051 | 5.6 | 20 | -90.1113 | 975.25 |
| 90 | 54.83051 | 11 | 55.5 | -90.1107 | 995.25 |
| 91 | 57.39063 | 8.3 | 7.8 | -88.9005 | 875.25 |
| 92 | 57.39063 | 9.4 | 7.2 | -88.9003 | 895.25 |
| 93 | 57.39063 | 8.2 | 27.6 | -88.9001 | 915.25 |
| 94 | 57.39063 | 13.2 | 14.8 | -88.9003 | 935.25 |
| 95 | 57.39063 | 9.9 | 51.2 | -88.9006 | 955.25 |
| 96 | 58.02985 | 10.9 | 14.6 | -88.9 | 975.25 |
| 97 | 58.02985 | 5.1 | 37.7 | -88.8991 | 995.25 |
| 98 | 60.03077 | 9.9 | 9.6 | -87.7248 | 875.25 |
| 99 | 60.03077 | 5.7 | 18.7 | -87.7257 | 895.25 |
| 100 | 60.03077 | 6.9 | 13.4 | -87.7242 | 915.25 |
| 101 | 60.03077 | 15 | 10.5 | -87.7249 | 935.25 |
| 102 | 60.03077 | 12.7 | 59.2 | -87.7245 | 955.25 |
| 103 | 60.03077 | 12.2 | 33.4 | -87.7247 | 975.25 |
| 104 | 60.03077 | 8.9 | 33.8 | -87.7245 | 995.25 |
| 105 | 66.86667 | 12.7 | 15 | -86.4704 | 875.25 |
| 106 | 66.86667 | 11.4 | 6.6 | -86.4712 | 895.25 |
| 107 | 66.86667 | 5.8 | 18.2 | -86.4713 | 915.25 |
| 108 | 66.86667 | 14.5 | 11.7 | -86.4705 | 935.25 |
| 109 | 66.86667 | 10.7 | 54.6 | -86.4709 | 955.25 |
| 110 | 66.86667 | 13.4 | 23.7 | -86.4713 | 975.25 |
| 111 | 66.86667 | 9.3 | 69.1 | -86.4709 | 995.25 |
| 112 | 69.01786 | 8.6 | 6.4 | -85.25 | 875.25 |
| 113 | 69.01786 | 10.7 | 20.1 | -85.2502 | 895.25 |
| 114 | 69.01786 | 9 | 10.6 | -85.2499 | 915.25 |
| 115 | 69.01786 | 16 | 13.6 | -85.25 | 935.25 |
| 116 | 69.01786 | 5.6 | 19.1 | -85.251 | 955.25 |
| 117 | 69.01786 | 9.2 | 68.6 | -85.2507 | 975.25 |
| 118 | 69.01786 | 7.3 | 22.8 | -85.2501 | 995.25 |
| 119 | 29.62791 | 7.9 | 9.2 | -95.5716 | 865.25 |
| 120 | 29.62791 | 21.2 | 9.8 | -95.5717 | 885.25 |
| 121 | 29.62791 | 13.7 | 8 | -95.5719 | 905.25 |
| 122 | 29.62791 | 9.9 | 33.2 | -95.5718 | 925.25 |
| 123 | 29.62791 | 6.9 | 18.4 | -95.5716 | 945.25 |
| 124 | 29.62791 | 15.6 | 9 | -95.5719 | 965.25 |
| 125 | 29.62791 | 9.2 | 36.3 | -95.5716 | 985.25 |
| 126 | 29.62791 | 4.6 | 15.4 | -95.5718 | 1005.25 |
| 127 | 29.06522 | 8.5 | 5.6 | -94.7963 | 865.25 |
| 128 | 29.06522 | 23.7 | 16.8 | -94.7964 | 885.25 |
| 129 | 29.06522 | 8.1 | 10.7 | -94.7971 | 905.25 |
| 130 | 29.06522 | 11.8 | 6.6 | -94.7967 | 925.25 |
| 131 | 29.06522 | 19.4 | 32.8 | -94.7964 | 945.25 |
| 132 | 29.06522 | 5.5 | 40.4 | -94.7961 | 965.25 |
| 133 | 29.06522 | 10.8 | 26 | -94.7964 | 985.25 |
| 134 | 29.06522 | 4.3 | 12.6 | -94.7964 | 1005.25 |
| 135 | 28.97959 | 16.2 | 6.2 | -94.0161 | 865.25 |
| 136 | 28.97959 | 16.3 | 5.6 | -94.016 | 885.25 |
| 137 | 28.97959 | 15.4 | 11.4 | -94.0157 | 905.25 |
| 138 | 28.97959 | 8.3 | 29.2 | -94.0159 | 925.25 |
| 139 | 28.97959 | 5.7 | 35.4 | -94.0148 | 945.25 |
| 140 | 28.97959 | 4.9 | 13 | -94.0158 | 965.25 |
| **141** | **28.97959** | **15.7** | **31.4** | **-94.0163** | **985.25** |
| 142 | 28.97959 | 9 | 14.5 | -94.0163 | 1005.25 |
| 143 | 30.55172 | 23.7 | 4.6 | -93.2425 | 865.25 |
| 144 | 30.55172 | 18.7 | 10.1 | -93.2417 | 885.25 |
| 145 | 30.55172 | 6.7 | 9.9 | -93.2419 | 905.25 |
| 146 | 30.55172 | 14.5 | 26.2 | -93.242 | 925.25 |
| 147 | 30.55172 | 10.8 | 9.4 | -93.2426 | 945.25 |
| **148** | **30.55172** | **6.7** | **47.6** | **-93.2425** | **965.25** |
| 149 | 30.55172 | 5 | 30.3 | -93.2418 | 985.25 |
| 150 | 30.55172 | 9.2 | 20.4 | -93.2424 | 1005.25 |
| 151 | 30.48276 | 27 | 12.7 | -92.4729 | 865.25 |
| 152 | 30.48276 | 12.4 | 7.5 | -92.4726 | 885.25 |
| 153 | 30.48276 | 6.2 | 19.2 | -92.4723 | 905.25 |
| 154 | 30.48276 | 8.8 | 12.3 | -92.4727 | 925.25 |
| 155 | 30.48276 | 18.4 | 8 | -92.4726 | 945.25 |
| 156 | 30.48276 | 13.1 | 38.9 | -92.4725 | 965.25 |
| 157 | 30.48276 | 11.4 | 21 | -92.4725 | 985.25 |
| 158 | 30.48276 | 7.9 | 40 | -92.4724 | 1005.25 |
| 159 | 30.15 | 19.8 | 5.9 | -91.6849 | 865.25 |
| 160 | 30.15 | 16.8 | 20.2 | -91.6852 | 885.25 |
| 161 | 30.15 | 21 | 9.2 | -91.685 | 905.25 |
| 162 | 30.15 | 7.2 | 13.9 | -91.6848 | 925.25 |
| 163 | 30.15 | 14.5 | 7.4 | -91.6843 | 945.25 |
| **164** | **30.15** | **5.9** | **37.5** | **-91.6849** | **965.25** |
| **165** | **30.15** | **13.3** | **35** | **-91.6848** | **985.25** |
| 166 | 30.15 | 7.3 | 10.7 | -91.6844 | 1005.25 |
| 167 | 34.58333 | 13.9 | 13.3 | -90.8992 | 865.25 |
| 168 | 34.58333 | 9.1 | 5.3 | -90.899 | 885.25 |
| 169 | 34.58333 | 5.6 | 14.4 | -90.8989 | 905.25 |
| 170 | 34.58333 | 15.4 | 23.6 | -90.8992 | 925.25 |
| **171** | **34.58333** | **6.4** | **29.5** | **-90.8993** | **945.25** |
| **172** | **34.58333** | **9.1** | **32.5** | **-90.899** | **965.25** |
| 173 | 34.58333 | 11.1 | 72.2 | -90.8988 | 985.25 |
| 174 | 34.58333 | 8.5 | 16.4 | -90.8984 | 1005.25 |
| 175 | 34.83051 | 9.2 | 6.8 | -90.1173 | 865.25 |
| 176 | 34.83051 | 9.4 | 13.8 | -90.1173 | 885.25 |
| 177 | 34.83051 | 14.5 | 14 | -90.117 | 905.25 |
| **178** | **34.83051** | **7.7** | **39.1** | **-90.1168** | **925.25** |
| 179 | 34.83051 | 15 | 14.8 | -90.1171 | 945.25 |
| **180** | **34.83051** | **19.3** | **29.2** | **-90.1171** | **965.25** |
| **181** | **34.83051** | **9.4** | **27** | **-90.1167** | **985.25** |
| 182 | 34.83051 | 9.7 | 59.3 | -90.1174 | 1005.25 |
| 183 | 35.16129 | 11.9 | 9.5 | -89.3274 | 865.25 |
| 184 | 35.16129 | 13.2 | 15.1 | -89.3277 | 885.25 |
| 185 | 35.16129 | 7.5 | 7.7 | -89.327 | 905.25 |
| 186 | 35.16129 | 5.9 | 27.1 | -89.3273 | 925.25 |
| 187 | 35.16129 | 15.5 | 22.2 | -89.3271 | 945.25 |
| 188 | 35.16129 | 10.4 | 64.3 | -89.3273 | 965.25 |
| 189 | 35.16129 | 17.7 | 12.6 | -89.3277 | 985.25 |
| **190** | **35.16129** | **6.4** | **54.2** | **-89.3281** | **1005.25** |
| 191 | 37.71014 | 12.3 | 11.7 | -88.543 | 865.25 |
| 192 | 37.71014 | 7.6 | 17.5 | -88.5434 | 885.25 |
| 193 | 37.71014 | 16.5 | 13.6 | -88.5435 | 905.25 |
| 194 | 37.71014 | 13.2 | 7.7 | -88.5424 | 925.25 |
| 195 | 37.71014 | 7.6 | 19.1 | -88.5434 | 945.25 |
| 196 | 37.71014 | 8.1 | 59.3 | -88.5433 | 965.25 |
| 197 | 37.71014 | 15 | 24.9 | -88.5434 | 985.25 |
| 198 | 37.71014 | 11.5 | 48.1 | -88.5431 | 1005.25 |
| 199 | 40.03077 | 8.2 | 6.6 | -87.7427 | 865.25 |
| 200 | 40.03077 | 17.4 | 7.2 | -87.7434 | 885.25 |
| 201 | 40.03077 | 12.1 | 16.8 | -87.7433 | 905.25 |
| 202 | 40.03077 | 9.3 | 8.2 | -87.7439 | 925.25 |
| 203 | 40.03077 | 12.1 | 34.1 | -87.743 | 945.25 |
| 204 | 40.03077 | 5.2 | 16.3 | -87.7433 | 965.25 |
| 205 | 40.03077 | 9.9 | 66.6 | -87.743 | 985.25 |
| 206 | 40.03077 | 10.3 | 32.2 | -87.7432 | 1005.25 |
| 207 | 43.95082 | 6.6 | 5.5 | -86.9481 | 865.25 |
| 208 | 43.95082 | 11.7 | 11.6 | -86.9488 | 885.25 |
| 209 | 43.95082 | 14 | 7.3 | -86.9486 | 905.25 |
| 210 | 43.95082 | 5.7 | 31.7 | -86.9491 | 925.25 |
| 211 | 43.95082 | 9.9 | 30.5 | -86.9481 | 945.25 |
| 212 | 43.95082 | 11.6 | 36.2 | -86.9482 | 965.25 |
| 213 | 43.95082 | 6.4 | 15.3 | -86.9487 | 985.25 |
| 214 | 43.95082 | 15.2 | 18.4 | -86.9481 | 1005.25 |
| 215 | 47.26667 | 13.5 | 8.2 | -86.1468 | 865.25 |
| 216 | 47.26667 | 5.5 | 9 | -86.1468 | 885.25 |
| 217 | 47.26667 | 10.8 | 7 | -86.1458 | 905.25 |
| 218 | 47.26667 | 12.5 | 21.3 | -86.147 | 925.25 |
| 219 | 47.26667 | 8 | 44.1 | -86.1462 | 945.25 |
| 220 | 47.26667 | 13.5 | 18 | -86.1468 | 965.25 |
| 221 | 47.26667 | 7.2 | 42.1 | -86.1464 | 985.25 |
| 222 | 47.26667 | 6 | 13.5 | -86.1458 | 1005.25 |
| 223 | 9.627907 | 18.6 | 8 | -95.6178 | 875.25 |
| 224 | 9.627907 | 9.8 | 10.4 | -95.6183 | 895.25 |
| 225 | 9.627907 | 17.1 | 20 | -95.6178 | 915.25 |
| 226 | 9.627907 | 5.3 | 35.9 | -95.6178 | 935.25 |
| 227 | 9.627907 | 15.5 | 7.8 | -95.6181 | 955.25 |
| 228 | 9.627907 | 5.1 | 17.2 | -95.6176 | 975.25 |
| 229 | 9.627907 | 10.7 | 58.1 | -95.6175 | 995.25 |
| 230 | 9.065217 | 15.3 | 6.6 | -94.9837 | 875.25 |
| 231 | 9.065217 | 5.2 | 19.3 | -94.9838 | 895.25 |
| 232 | 9.065217 | 11.1 | 6.3 | -94.9839 | 915.25 |
| 233 | 9.065217 | 20.1 | 9.9 | -94.9838 | 935.25 |
| 234 | 9.065217 | 12 | 31.2 | -94.9831 | 955.25 |
| 235 | 9.065217 | 7.4 | 27.8 | -94.9835 | 975.25 |
| 236 | 9.065217 | 7 | 18.8 | -94.9836 | 995.25 |
| 237 | 9.469388 | 15.8 | 15.7 | -94.3494 | 875.25 |
| 238 | 9.469388 | 6.9 | 10.7 | -94.349 | 895.25 |
| 239 | 9.469388 | 18.5 | 17.1 | -94.3493 | 915.25 |
| 240 | 9.469388 | 12.9 | 8.2 | -94.3491 | 935.25 |
| 241 | 9.469388 | 18.7 | 20.8 | -94.3494 | 955.25 |
| **242** | **9.469388** | **6.4** | **56.1** | **-94.349** | **975.25** |
| 243 | 9.469388 | 6.2 | 26.9 | -94.3489 | 995.25 |
| 244 | 11.03846 | 17.3 | 10.6 | -93.7128 | 875.25 |
| 245 | 11.03846 | 14 | 5.5 | -93.7131 | 895.25 |
| 246 | 11.03846 | 6 | 15.9 | -93.7127 | 915.25 |
| 247 | 11.03846 | 18.1 | 12.3 | -93.7127 | 935.25 |
| **248** | **11.03846** | **10.2** | **28** | **-93.7128** | **955.25** |
| 249 | 11.03846 | 8.1 | 18.1 | -93.713 | 975.25 |
| 250 | 11.03846 | 7.6 | 50.9 | -93.7131 | 995.25 |
| 251 | 11.24138 | 23 | 7 | -93.0786 | 875.25 |
| 252 | 11.24138 | 11.8 | 12.1 | -93.0777 | 895.25 |
| **253** | **11.24138** | **15** | **34** | **-93.0781** | **915.25** |
| 254 | 11.24138 | 6.3 | 7.6 | -93.0782 | 935.25 |
| **255** | **11.24138** | **5.1** | **21.6** | **-93.0782** | **955.25** |
| 256 | 11.24138 | 10.6 | 20.9 | -93.0784 | 975.25 |
| **257** | **11.24138** | **11.3** | **35.1** | **-93.0782** | **995.25** |
| 258 | 10.48276 | 7.4 | 8.4 | -92.4393 | 875.25 |
| 259 | 10.48276 | 18.9 | 8.2 | -92.4392 | 895.25 |
| 260 | 10.48276 | 21.5 | 28.9 | -92.4393 | 915.25 |
| 261 | 10.48276 | 10.8 | 17.4 | -92.4393 | 935.25 |
| 262 | 10.48276 | 4.7 | 18.4 | -92.4391 | 955.25 |
| **263** | **10.48276** | **11.2** | **38.6** | **-92.4391** | **975.25** |
| 264 | 10.48276 | 5.3 | 9.3 | -92.4389 | 995.25 |
| 265 | 10.10169 | 11 | 11.5 | -91.7978 | 875.25 |
| 266 | 10.10169 | 6 | 7.6 | -91.7972 | 895.25 |
| 267 | 10.10169 | 5 | 26.5 | -91.797 | 915.25 |
| **268** | **10.10169** | **10.2** | **29.5** | **-91.797** | **935.25** |
| 269 | 10.10169 | 7.8 | 10 | -91.7975 | 955.25 |
| **270** | **10.10169** | **8.5** | **22.8** | **-91.7972** | **975.25** |
| 271 | 10.10169 | 4.9 | 42 | -91.7972 | 995.25 |
| 272 | 12.82759 | 14.8 | 11.8 | -91.1511 | 875.25 |
| 273 | 12.82759 | 13.6 | 14.9 | -91.1512 | 895.25 |
| 274 | 12.82759 | 7.9 | 14.9 | -91.1508 | 915.25 |
| 275 | 12.82759 | 5.6 | 28.5 | -91.151 | 935.25 |
| **276** | **12.82759** | **10.4** | **46.1** | **-91.151** | **955.25** |
| **277** | **12.82759** | **7.6** | **26.7** | **-91.1507** | **975.25** |
| **278** | **12.82759** | **17** | **31.4** | **-91.151** | **995.25** |
| 279 | 12.50909 | 10.3 | 4.9 | -90.515 | 875.25 |
| 280 | 10.10169 | 8 | 17.4 | -91.7973 | 895.25 |
| **281** | **10.10169** | **15.9** | **36.6** | **-91.7974** | **915.25** |
| 282 | 10.10169 | 6.5 | 21.6 | -91.7977 | 935.25 |
| **283** | **10.10169** | **17.7** | **16.7** | **-91.7975** | **955.25** |
| 284 | 10.10169 | 14.4 | 21.8 | -91.7973 | 975.25 |
| **285** | **10.10169** | **9.8** | **43.6** | **-91.7971** | **995.25** |
| 286 | 14.77193 | 16.3 | 7.3 | -89.869 | 875.25 |
| 287 | 14.77193 | 8.6 | 8.5 | -89.8692 | 895.25 |
| **288** | **14.77193** | **5.3** | **25.5** | **-89.8691** | **915.25** |
| **289** | **14.77193** | **8.8** | **23.9** | **-89.8695** | **935.25** |
| **290** | **14.77193** | **11.3** | **44** | **-89.869** | **955.25** |
| 291 | 14.77193 | 8.3 | 16.4 | -89.8694 | 975.25 |
| **292** | **14.77193** | **13.2** | **53.1** | **-89.8689** | **995.25** |
| 293 | 16.70313 | 10.6 | 9.1 | -89.2264 | 875.25 |
| 294 | 16.70313 | 12.9 | 7.9 | -89.2268 | 895.25 |
| 295 | 16.70313 | 17.8 | 16.5 | -89.2269 | 915.25 |
| **296** | **16.70313** | **5.8** | **19.4** | **-89.2265** | **935.25** |
| **297** | **16.70313** | **6.7** | **40.4** | **-89.227** | **955.25** |
| **298** | **16.70313** | **15** | **37.1** | **-89.2265** | **975.25** |
| **299** | **16.70313** | **6.8** | **15.9** | **-89.2262** | **995.25** |
| 300 | 17.71014 | 5.9 | 16.6 | -88.5732 | 875.25 |
| 301 | 17.71014 | 16.4 | 12.5 | -88.5734 | 895.25 |
| 302 | 17.71014 | 10.2 | 8.4 | -88.5733 | 915.25 |
| 303 | 17.71014 | 9.9 | 13 | -88.5729 | 935.25 |
| **304** | **17.71014** | **6.5** | **23.3** | **-88.573** | **955.25** |
| 305 | 17.71014 | 13 | 59.3 | -88.5727 | 975.25 |
| 306 | 17.71014 | 10.1 | 17.8 | -88.5729 | 995.25 |
| 307 | 20.14925 | 14.4 | 5.2 | -87.9329 | 875.25 |
| 308 | 20.14925 | 8.3 | 21.9 | -87.9333 | 895.25 |
| 309 | 20.14925 | 15.4 | 15.4 | -87.9336 | 915.25 |
| 310 | 20.14925 | 19 | 33 | -87.933 | 935.25 |
| 311 | 20.14925 | 7.4 | 10.7 | -87.9331 | 955.25 |
| **312** | **20.14925** | **7.2** | **31** | **-87.9329** | **975.25** |
| 313 | 20.14925 | 15.9 | 24.8 | -87.9332 | 995.25 |
| 314 | 23.04762 | 7.1 | 5.6 | -87.2823 | 875.25 |
| 315 | 23.04762 | 17 | 15.4 | -87.2825 | 895.25 |
| 316 | 23.04762 | 7.6 | 22.9 | -87.2822 | 915.25 |
| 317 | 23.04762 | 17.3 | 25.6 | -87.2828 | 935.25 |
| 318 | 23.04762 | 8.9 | 14.5 | -87.2822 | 955.25 |
| 319 | 23.04762 | 10 | 63.3 | -87.2824 | 975.25 |
| 320 | 23.04762 | 10.4 | 30.3 | -87.2825 | 995.25 |
| 321 | 23.73438 | 13.9 | 8.2 | -86.6288 | 875.25 |
| 322 | 23.73438 | 7.7 | 14.5 | -86.6291 | 895.25 |
| 323 | 23.73438 | 9.9 | 11.6 | -86.6295 | 915.25 |
| 324 | 23.73438 | 12.1 | 39.5 | -86.6291 | 935.25 |
| 325 | 23.73438 | 7.6 | 29 | -86.6289 | 955.25 |
| 326 | 23.73438 | 5.7 | 41.5 | -86.6292 | 975.25 |
| 327 | 23.73438 | 12.4 | 14 | -86.6289 | 995.25 |
| 328 | -10.3721 | 5.4 | 7 | -95.54 | 865.25 |
| 329 | -10.3721 | 19.5 | 8.5 | -95.54 | 885.25 |
| 330 | -10.3721 | 9.2 | 6.2 | -95.54 | 905.25 |
| 331 | -10.3721 | 9.6 | 30.4 | -95.54 | 925.25 |
| 332 | -10.3721 | 12.8 | 23.1 | -95.54 | 945.25 |
| 333 | -10.3721 | 8.8 | 15.4 | -95.54 | 965.25 |
| 334 | -10.3721 | 7.8 | 37.7 | -95.54 | 985.25 |
| 335 | -10.3721 | 5 | 21.3 | -95.54 | 1005.25 |
| 336 | -10.9348 | 10.5 | 11.3 | -94.93 | 865.25 |
| 337 | -10.9348 | 22.3 | 6 | -94.93 | 885.25 |
| 338 | -10.9348 | 23.9 | 15.8 | -94.93 | 905.25 |
| 339 | -10.9348 | 13.7 | 22.1 | -94.93 | 925.25 |
| 340 | -10.9348 | 7.3 | 12.7 | -94.93 | 945.25 |
| 341 | -10.9348 | 5.7 | 43.7 | -94.93 | 965.25 |
| 342 | -10.9348 | 8.3 | 22.9 | -94.93 | 985.25 |
| 343 | -10.9348 | 6.2 | 27.6 | -94.93 | 1005.25 |
| 344 | -10.5306 | 15.7 | 7.1 | -94.32 | 865.25 |
| 345 | -10.5306 | 20.3 | 11 | -94.32 | 885.25 |
| 346 | -10.5306 | 19.1 | 8.4 | -94.32 | 905.25 |
| **347** | **-10.5306** | **11.1** | **36.7** | **-94.32** | **925.25** |
| 348 | -10.5306 | 9.1 | 11.4 | -94.32 | 945.25 |
| **349** | **-10.5306** | **14.4** | **33.6** | **-94.32** | **965.25** |
| 350 | -10.5306 | 4.9 | 32.6 | -94.32 | 985.25 |
| 351 | -10.5306 | 9.4 | 25.5 | -94.32 | 1005.25 |
| 352 | -8.96154 | 21.5 | 4.3 | -93.71 | 865.25 |
| 353 | -8.96154 | 6.7 | 10.4 | -93.71 | 885.25 |
| 354 | -8.96154 | 12.9 | 10.3 | -93.71 | 905.25 |
| **355** | **-8.96154** | **19.7** | **15.6** | **-93.71** | **925.25** |
| 356 | -8.96154 | 9.6 | 13.4 | -93.71 | 945.25 |
| 357 | -8.96154 | 7.2 | 52.5 | -93.71 | 965.25 |
| **358** | **-8.96154** | **8.7** | **40.6** | **-93.71** | **985.25** |
| 359 | -8.96154 | 6.6 | 17.4 | -93.71 | 1005.25 |
| 360 | -8.75862 | 19.1 | 9.7 | -93.1 | 865.25 |
| 361 | -8.75862 | 25.4 | 7 | -93.1 | 885.25 |
| 362 | -8.75862 | 20 | 17.4 | -93.1 | 905.25 |
| 363 | -8.75862 | 5.5 | 11.2 | -93.1 | 925.25 |
| **364** | **-8.75862** | **8.3** | **40** | **-93.1** | **945.25** |
| **365** | **-8.75862** | **6.3** | **30.2** | **-93.1** | **965.25** |
| 366 | -8.75862 | 6.6 | 9.8 | -93.1 | 985.25 |
| **367** | **-8.75862** | **11.2** | **31** | **-93.1** | **1005.25** |
| 368 | -9.51724 | 7 | 7.3 | -92.49 | 865.25 |
| 369 | -9.51724 | 14.8 | 6.7 | -92.49 | 885.25 |
| 370 | -9.51724 | 17.7 | 22 | -92.49 | 905.25 |
| 371 | -9.51724 | 20.8 | 14.5 | -92.49 | 925.25 |
| **372** | **-9.51724** | **11.1** | **31.6** | **-92.49** | **945.25** |
| **373** | **-9.51724** | **7.6** | **19.7** | **-92.49** | **965.25** |
| **374** | **-9.51724** | **5.2** | **49.1** | **-92.49** | **985.25** |
| 375 | -9.51724 | 5.5 | 9.7 | -92.49 | 1005.25 |
| 376 | -9.77966 | 17.8 | 12.1 | -91.88 | 865.25 |
| 377 | -9.77966 | 12.8 | 8.3 | -91.88 | 885.25 |
| 378 | -9.77966 | 5.3 | 22.8 | -91.88 | 905.25 |
| 379 | -9.77966 | 17.9 | 7.1 | -91.88 | 925.25 |
| 380 | -9.77966 | 11.4 | 21.4 | -91.88 | 945.25 |
| 381 | -9.77966 | 4.6 | 10.6 | -91.88 | 965.25 |
| **382** | **-9.77966** | **5.8** | **43.6** | **-91.88** | **985.25** |
| 383 | -9.77966 | 8.7 | 34.6 | -91.88 | 1005.25 |
| 384 | -7.50877 | 17.2 | 10.6 | -91.27 | 865.25 |
| 385 | -7.50877 | 8.7 | 13.4 | -91.27 | 885.25 |
| 386 | -7.50877 | 7.2 | 11.1 | -91.27 | 905.25 |
| 387 | -7.50877 | 12.9 | 12.9 | -91.27 | 925.25 |
| **388** | **-7.50877** | **10.2** | **46.5** | **-91.27** | **945.25** |
| 389 | -7.50877 | 13.9 | 26.1 | -91.27 | 965.25 |
| 390 | -7.50877 | 7.6 | 18.1 | -91.27 | 985.25 |
| **391** | **-7.50877** | **8.1** | **41.5** | **-91.27** | **1005.25** |
| 392 | -6.81356 | 8.9 | 4.5 | -90.66 | 865.25 |
| 393 | -6.81356 | 10.7 | 21.7 | -90.66 | 885.25 |
| **394** | **-6.81356** | **5.9** | **25.9** | **-90.66** | **905.25** |
| **395** | **-6.81356** | **11.4** | **18.6** | **-90.66** | **925.25** |
| **396** | **-6.81356** | **5.9** | **24.7** | **-90.66** | **945.25** |
| **397** | **-6.81356** | **11.3** | **18.8** | **-90.66** | **965.25** |
| 398 | -6.81356 | 7 | 10.7 | -90.66 | 985.25 |
| 399 | -6.81356 | 8.3 | 70.5 | -90.66 | 1005.25 |
| 400 | -5.22807 | 13.1 | 7.5 | -90.05 | 865.25 |
| 401 | -5.22807 | 14 | 10.7 | -90.05 | 885.25 |
| 402 | -5.22807 | 18.4 | 23.8 | -90.05 | 905.25 |
| 403 | -5.22807 | 10.8 | 10.7 | -90.05 | 925.25 |
| 404 | -5.22807 | 5.2 | 16.9 | -90.05 | 945.25 |
| **405** | **-5.22807** | **7.4** | **28.1** | **-90.05** | **965.25** |
| 406 | -5.22807 | 9 | 62.4 | -90.05 | 985.25 |
| 407 | -5.22807 | 13.7 | 26.5 | -90.05 | 1005.25 |
| 408 | -2.91667 | 12.7 | 4.9 | -89.44 | 865.25 |
| 409 | -2.91667 | 8.2 | 18.3 | -89.44 | 885.25 |
| 410 | -2.91667 | 6.4 | 9.5 | -89.44 | 905.25 |
| 411 | -2.91667 | 15 | 16.8 | -89.44 | 925.25 |
| 412 | -2.91667 | 12.4 | 8.7 | -89.44 | 945.25 |
| **413** | **-2.91667** | **11.9** | **49.9** | **-89.44** | **965.25** |
| 414 | -2.91667 | 13.8 | 17.2 | -89.44 | 985.25 |
| 415 | -2.91667 | 5.7 | 46.3 | -89.44 | 1005.25 |
| 416 | -2.02899 | 5.7 | 4.8 | -88.83 | 865.25 |
| 417 | -2.02899 | 10 | 9.6 | -88.83 | 885.25 |
| 418 | -2.02899 | 10.2 | 31.3 | -88.83 | 905.25 |
| 419 | -2.02899 | 14 | 24.4 | -88.83 | 925.25 |
| 420 | -2.02899 | 5.5 | 25.6 | -88.83 | 945.25 |
| 421 | -2.02899 | 7.9 | 9.8 | -88.83 | 965.25 |
| **422** | **-2.02899** | **6** | **39.1** | **-88.83** | **985.25** |
| 423 | -2.02899 | 12.7 | 49.9 | -88.83 | 1005.25 |
| 424 | -1.38235 | 14.7 | 8 | -88.22 | 865.25 |
| 425 | -1.38235 | 5.8 | 12.3 | -88.22 | 885.25 |
| 426 | -1.38235 | 12.5 | 13.1 | -88.22 | 905.25 |
| 427 | -1.38235 | 9.1 | 15.1 | -88.22 | 925.25 |
| 428 | -1.38235 | 10.5 | 16.2 | -88.22 | 945.25 |
| 429 | -1.38235 | 15 | 34.9 | -88.22 | 965.25 |
| 430 | -1.38235 | 8 | 53.5 | -88.22 | 985.25 |
| 431 | -1.38235 | 7 | 22.4 | -88.22 | 1005.25 |
| 432 | 0.030769 | 7.6 | 5.7 | -87.61 | 865.25 |
| 433 | 0.030769 | 15.8 | 7.7 | -87.61 | 885.25 |
| 434 | 0.030769 | 9 | 20.5 | -87.61 | 905.25 |
| 435 | 0.030769 | 16 | 16.2 | -87.61 | 925.25 |
| 436 | 0.030769 | 13.2 | 12 | -87.61 | 945.25 |
| 437 | 0.030769 | 16.2 | 55.6 | -87.61 | 965.25 |
| 438 | 0.030769 | 10.5 | 28.1 | -87.61 | 985.25 |
| **439** | **0.030769** | **6.7** | **38.6** | **-87.61** | **1005.25** |
| **440** | **3.746032** | **10.2** | **14.2** | **-87** | **865.25** |
| 441 | 3.746032 | 8.5 | 5.8 | -87 | 885.25 |
| 442 | 3.746032 | 15 | 12.3 | -87 | 905.25 |
| 443 | 3.746032 | 18.7 | 19.3 | -87 | 925.25 |
| 444 | 3.746032 | 8.6 | 19.9 | -87 | 945.25 |
| 445 | 3.746032 | 6.5 | 21.4 | -87 | 965.25 |
| 446 | 3.746032 | 12.1 | 56 | -87 | 985.25 |
| 447 | 3.746032 | 10.5 | 19.3 | -87 | 1005.25 |
| 448 | -30.3721 | 16.8 | 6.7 | -95.6029 | 875.25 |
| 449 | -30.3721 | 5.5 | 21 | -95.6032 | 895.25 |
| 450 | -30.3721 | 10.5 | 6.7 | -95.6027 | 915.25 |
| 451 | -30.3721 | 21.6 | 11.1 | -95.6035 | 935.25 |
| 452 | -30.3721 | 12.4 | 33.5 | -95.603 | 955.25 |
| 453 | -30.3721 | 7.8 | 29.9 | -95.6029 | 975.25 |
| 454 | -30.3721 | 7.2 | 19.8 | -95.6029 | 995.25 |
| 455 | -30.9348 | 12 | 5.5 | -94.9587 | 875.25 |
| 456 | -30.9348 | 10.4 | 5.8 | -94.9584 | 895.25 |
| 457 | -30.9348 | 11.4 | 23.7 | -94.9587 | 915.25 |
| 458 | -30.9348 | 15.5 | 16.1 | -94.9584 | 935.25 |
| 459 | -30.9348 | 8.6 | 24.2 | -94.9584 | 955.25 |
| 460 | -30.9348 | 6.7 | 8.6 | -94.9586 | 975.25 |
| 461 | -30.9348 | 4.4 | 21.8 | -94.9581 | 995.25 |
| 462 | -30.5306 | 12.4 | 5.8 | -94.3234 | 875.25 |
| 463 | -30.5306 | 12.1 | 6 | -94.3233 | 895.25 |
| 464 | -30.5306 | 11.8 | 24.6 | -94.3235 | 915.25 |
| 465 | -30.5306 | 16.6 | 16.7 | -94.3231 | 935.25 |
| **466** | **-30.5306** | **9.1** | **26.1** | **-94.3232** | **955.25** |
| 467 | -30.5306 | 7 | 9.4 | -94.3238 | 975.25 |
| 468 | -30.5306 | 4.6 | 23.8 | -94.3233 | 995.25 |
| 469 | -28.9615 | 8 | 8.9 | -93.6883 | 875.25 |
| 470 | -28.9615 | 20.6 | 8.8 | -93.6879 | 895.25 |
| **471** | **-28.9615** | **19.3** | **30.3** | **-93.6881** | **915.25** |
| 472 | -28.9615 | 11.4 | 18.7 | -93.6883 | 935.25 |
| 473 | -28.9615 | 4.9 | 19.9 | -93.6881 | 955.25 |
| **474** | **-28.9615** | **9** | **40** | **-93.6876** | **975.25** |
| 475 | -28.9615 | 5.5 | 10.3 | -93.6877 | 995.25 |
| 476 | -28.7586 | 6.2 | 9.8 | -93.0572 | 875.25 |
| 477 | -28.7586 | 24.7 | 13.7 | -93.0568 | 895.25 |
| 478 | -28.7586 | 13.7 | 7.1 | -93.0573 | 915.25 |
| **479** | **-28.7586** | **8.5** | **30.6** | **-93.0574** | **935.25** |
| **480** | **-28.7586** | **13.1** | **15.2** | **-93.0567** | **955.25** |
| 481 | -28.7586 | 4.5 | 13.7 | -93.0577 | 975.25 |
| **482** | **-28.7586** | **8** | **45.2** | **-93.0571** | **995.25** |
| 483 | -29.5172 | 9 | 4.7 | -92.4166 | 875.25 |
| 484 | -29.5172 | 7.4 | 26.2 | -92.4166 | 895.25 |
| 485 | -29.5172 | 23.1 | 17.6 | -92.4166 | 915.25 |
| **486** | **-29.5172** | **7** | **27.5** | **-92.4167** | **935.25** |
| 487 | -29.5172 | 6 | 11.4 | -92.4168 | 955.25 |
| **488** | **-29.5172** | **15.6** | **12** | **-92.4165** | **975.25** |
| 489 | -29.5172 | 8.6 | 74.8 | -92.4166 | 995.25 |
| 490 | -29.8983 | 11.7 | 10.1 | -91.7894 | 875.25 |
| 491 | -29.8983 | 19.7 | 11.4 | -91.7894 | 895.25 |
| 492 | -29.8983 | 7.1 | 8 | -91.7895 | 915.25 |
| 493 | -29.8983 | 6 | 18 | -91.7897 | 935.25 |
| **494** | **-29.8983** | **11.7** | **36** | **-91.7894** | **955.25** |
| 495 | -29.8983 | 16.4 | 12.9 | -91.7894 | 975.25 |
| 496 | -29.8983 | 7.8 | 61.1 | -91.7894 | 995.25 |
| 497 | -27.1724 | 6.5 | 12.5 | -91.1497 | 875.25 |
| 498 | -27.1724 | 18.2 | 9.1 | -91.1499 | 895.25 |
| 499 | -27.1724 | 20.4 | 31.9 | -91.15 | 915.25 |
| **500** | **-27.1724** | **8.3** | **26.6** | **-91.1502** | **935.25** |
| 501 | -27.1724 | 8.1 | 8.5 | -91.1499 | 955.25 |
| 502 | -27.1724 | 5 | 34.6 | -91.1502 | 975.25 |
| 503 | -27.1724 | 11.6 | 39.1 | -91.1499 | 995.25 |
| 504 | -27.4909 | 17.9 | 9.3 | -90.5184 | 875.25 |
| **505** | **-27.4909** | **6.3** | **15.9** | **-90.5178** | **895.25** |
| 506 | -27.4909 | 12.1 | 18.8 | -90.5182 | 915.25 |
| 507 | -27.4909 | 7.8 | 8.7 | -90.5185 | 935.25 |
| 508 | -27.4909 | 9.3 | 38.8 | -90.5184 | 955.25 |
| **509** | **-27.4909** | **4.8** | **24.7** | **-90.5184** | **975.25** |
| 510 | -27.4909 | 13.8 | 36.4 | -90.5181 | 995.25 |
| 511 | -25.2281 | 5.6 | 7.2 | -89.8916 | 875.25 |
| 512 | -25.2281 | 11.1 | 16.9 | -89.8916 | 895.25 |
| 513 | -25.2281 | 10.8 | 11.1 | -89.8916 | 915.25 |
| 514 | -25.2281 | 13.6 | 20.8 | -89.8916 | 935.25 |
| 515 | -25.2281 | 14 | 17.5 | -89.8916 | 955.25 |
| **516** | **-25.2281** | **8.7** | **53.3** | **-89.8917** | **975.25** |
| 517 | -25.2281 | 5.7 | 25.9 | -89.8917 | 995.25 |
| 518 | -23.2969 | 9.3 | 8.6 | -89.26 | 875.25 |
| 519 | -23.2969 | 6.6 | 24.4 | -89.2601 | 895.25 |
| 520 | -23.2969 | 6.6 | 12.5 | -89.2601 | 915.25 |
| 521 | -23.2969 | 12.5 | 31.7 | -89.26 | 935.25 |
| 522 | -23.2969 | 16.1 | 13.7 | -89.2602 | 955.25 |
| 523 | -23.2969 | 10.3 | 25.7 | -89.2597 | 975.25 |
| **524** | **-23.2969** | **6.4** | **40.5** | **-89.2599** | **995.25** |
| 525 | -22.3971 | 7.7 | 6 | -88.6229 | 875.25 |
| 526 | -22.3971 | 10.1 | 10 | -88.623 | 895.25 |
| 527 | -22.3971 | 8.4 | 22.1 | -88.6226 | 915.25 |
| **528** | **-22.3971** | **7.3** | **37.5** | **-88.6226** | **935.25** |
| **529** | **-22.3971** | **15** | **25.1** | **-88.6229** | **955.25** |
| **530** | **-22.3971** | **9.5** | **11.1** | **-88.6229** | **975.25** |
| 531 | -22.3971 | 5.8 | 48.9 | -88.6228 | 995.25 |
| 532 | -19.8507 | 11.3 | 10.9 | -87.9992 | 875.25 |
| 533 | -19.8507 | 12.5 | 11.1 | -87.9995 | 895.25 |
| 534 | -19.8507 | 14 | 9.3 | -87.9994 | 915.25 |
| 535 | -19.8507 | 6.8 | 34.4 | -87.9992 | 935.25 |
| 536 | -19.8507 | 20 | 30.1 | -87.9993 | 955.25 |
| 537 | -19.8507 | 5.3 | 19.1 | -87.999 | 975.25 |
| 538 | -19.8507 | 12.8 | 29.2 | -87.9994 | 995.25 |
| 539 | -17.5 | 19.3 | 10.3 | -87.3666 | 875.25 |
| 540 | -17.5 | 7.1 | 14 | -87.3669 | 895.25 |
| 541 | -17.5 | 13.2 | 12 | -87.3672 | 915.25 |
| 542 | -17.5 | 11.1 | 23.1 | -87.3669 | 935.25 |
| 543 | -17.5 | 7.1 | 32.3 | -87.3669 | 955.25 |
| 544 | -17.5 | 18.5 | 43.1 | -87.3673 | 975.25 |
| 545 | -17.5 | 8.2 | 13.1 | -87.3671 | 995.25 |
| 546 | -49.6939 | 22.5 | 5.2 | -94.1419 | 865.25 |
| 547 | -49.6939 | 7.9 | 13 | -94.1421 | 885.25 |
| 548 | -49.6939 | 9.8 | 12.7 | -94.1424 | 905.25 |
| 549 | -49.6939 | 22.4 | 20 | -94.1412 | 925.25 |
| **550** | **-49.6939** | **8.9** | **53.7** | **-94.1425** | **945.25** |
| 551 | -49.6939 | 10.1 | 8.2 | -94.1423 | 965.25 |
| 552 | -49.6939 | 5.4 | 16.2 | -94.1432 | 985.25 |
| 553 | -49.6939 | 10 | 37.2 | -94.1426 | 1005.25 |
| 554 | -49.4483 | 6 | 7.8 | -93.3481 | 865.25 |
| 555 | -49.4483 | 11 | 12.6 | -93.3489 | 885.25 |
| 556 | -49.4483 | 22.2 | 18.5 | -93.3481 | 905.25 |
| 557 | -49.4483 | 6.4 | 18 | -93.3482 | 925.25 |
| 558 | -49.4483 | 13.6 | 10.7 | -93.348 | 945.25 |
| 559 | -49.4483 | 18 | 12.2 | -93.3483 | 965.25 |
| **560** | **-49.4483** | **6.8** | **29.2** | **-93.3481** | **985.25** |
| 561 | -49.4483 | 7.5 | 76.3 | -93.3485 | 1005.25 |
| 562 | -48.9483 | 14.3 | 8.4 | -92.5574 | 865.25 |
| 563 | -48.9483 | 13.6 | 6.5 | -92.5576 | 885.25 |
| 564 | -48.9483 | 17 | 24.8 | -92.5572 | 905.25 |
| 565 | -48.9483 | 6.2 | 20.6 | -92.5577 | 925.25 |
| 566 | -48.9483 | 20.8 | 23.9 | -92.5573 | 945.25 |
| **567** | **-48.9483** | **10.7** | **14.6** | **-92.5574** | **965.25** |
| 568 | -48.9483 | 6.2 | 14.4 | -92.5577 | 985.25 |
| 569 | -48.9483 | 4.5 | 52 | -92.5578 | 1005.25 |
| 570 | -49.8983 | 25.1 | 10 | -91.7639 | 865.25 |
| 571 | -49.8983 | 9.7 | 11.3 | -91.7647 | 885.25 |
| 572 | -49.8983 | 11.1 | 6.6 | -91.7654 | 905.25 |
| **573** | **-49.8983** | **5.2** | **28.1** | **-91.7642** | **925.25** |
| 574 | -49.8983 | 5 | 14.1 | -91.7649 | 945.25 |
| **575** | **-49.8983** | **12.7** | **17.1** | **-91.7652** | **965.25** |
| **576** | **-49.8983** | **14.4** | **45.4** | **-91.7643** | **985.25** |
| 577 | -49.8983 | 5.9 | 28.7 | -91.7641 | 1005.25 |
| 578 | -45.3509 | 7.2 | 6.5 | -90.9759 | 865.25 |
| 579 | -45.3509 | 6.1 | 16.2 | -90.9762 | 885.25 |
| 580 | -45.3509 | 9.6 | 11.9 | -90.9762 | 905.25 |
| 581 | -45.3509 | 16.6 | 22.8 | -90.976 | 925.25 |
| 582 | -45.3509 | 11.7 | 10 | -90.9763 | 945.25 |
| **583** | **-45.3509** | **9.6** | **45.6** | **-90.9762** | **965.25** |
| **584** | **-45.3509** | **7.4** | **23.9** | **-90.9758** | **985.25** |
| 585 | -45.3509 | 11.9 | 24.4 | -90.9763 | 1005.25 |
| 586 | -45.1695 | 16.7 | 5.3 | -90.1883 | 865.25 |
| 587 | -45.1695 | 6.4 | 19.2 | -90.188 | 885.25 |
| 588 | -45.1695 | 8.4 | 8.8 | -90.1883 | 905.25 |
| 589 | -45.1695 | 10.5 | 17.4 | -90.1882 | 925.25 |
| **590** | **-45.1695** | **16.1** | **28.5** | **-90.1877** | **945.25** |
| 591 | -45.1695 | 7 | 24.2 | -90.1884 | 965.25 |
| 592 | -45.1695 | 16.6 | 47.1 | -90.1883 | 985.25 |
| 593 | -45.1695 | 14.4 | 11.6 | -90.1884 | 1005.25 |
| 594 | -42.9167 | 15.2 | 8.6 | -89.4177 | 865.25 |
| 595 | -42.9167 | 12 | 8.7 | -89.4178 | 885.25 |
| 596 | -42.9167 | 7 | 21.2 | -89.4181 | 905.25 |
| 597 | -42.9167 | 7 | 8.7 | -89.4181 | 925.25 |
| 598 | -42.9167 | 6.6 | 20.6 | -89.418 | 945.25 |
| 599 | -42.9167 | 11 | 42 | -89.418 | 965.25 |
| 600 | -42.9167 | 12.5 | 20 | -89.4176 | 985.25 |
| 601 | -42.9167 | 7.1 | 29.9 | -89.4179 | 1005.25 |
| 602 | -42.3971 | 18.4 | 10.3 | -88.6134 | 865.25 |
| 603 | -42.3971 | 7 | 9.3 | -88.6128 | 885.25 |
| 604 | -42.3971 | 7.8 | 15.3 | -88.6131 | 905.25 |
| 605 | -42.3971 | 7.5 | 13.4 | -88.6129 | 925.25 |
| **606** | **-42.3971** | **14** | **41.9** | **-88.613** | **945.25** |
| 607 | -42.3971 | 12.3 | 13.8 | -88.6135 | 965.25 |
| 608 | -42.3971 | 12.9 | 33.8 | -88.6133 | 985.25 |
| **609** | **-42.3971** | **4.8** | **43** | **-88.6134** | **1005.25** |
| 610 | -40.9375 | 11.6 | 6.4 | -87.8442 | 865.25 |
| 611 | -40.9375 | 10.4 | 14.2 | -87.8443 | 885.25 |
| 612 | -40.9375 | 15.9 | 14.9 | -87.8442 | 905.25 |
| 613 | -40.9375 | 12.1 | 9.7 | -87.8445 | 925.25 |
| 614 | -40.9375 | 7.1 | 38.3 | -87.8444 | 945.25 |
| 615 | -40.9375 | 6.1 | 20.6 | -87.8444 | 965.25 |
| 616 | -40.9375 | 5.6 | 13.4 | -87.8444 | 985.25 |
| 617 | -40.9375 | 16.2 | 66 | -87.8445 | 1005.25 |
| 618 | -69.5172 | 20 | 12.2 | -92.4188 | 875.25 |
| 619 | -69.5172 | 14.5 | 6.4 | -92.419 | 895.25 |
| 620 | -69.5172 | 6.3 | 21.4 | -92.4192 | 915.25 |
| 621 | -69.5172 | 11.8 | 20.1 | -92.4196 | 935.25 |
| 622 | -69.5172 | 11 | 48.4 | -92.4191 | 955.25 |
| 623 | -69.5172 | 5.9 | 10.3 | -92.4184 | 975.25 |
| 624 | -69.5172 | 14.4 | 20.7 | -92.4193 | 995.25 |
| 625 | -67.5088 | 13.1 | 6.2 | -91.1804 | 875.25 |
| 626 | -67.5088 | 21.7 | 18 | -91.1802 | 895.25 |
| 627 | -67.5088 | 16.5 | 10.2 | -91.1806 | 915.25 |
| 628 | -67.5088 | 5.1 | 24.8 | -91.18 | 935.25 |
| 629 | -67.5088 | 6.9 | 42.1 | -91.1807 | 955.25 |
| 630 | -67.5088 | 13.9 | 32.2 | -91.1807 | 975.25 |
| 631 | -67.5088 | 6.6 | 12.1 | -91.1807 | 995.25 |
| 632 | -65.2281 | 26.2 | 5.1 | -89.9662 | 875.25 |
| 633 | -65.2281 | 8.9 | 23.1 | -89.9663 | 895.25 |
| 634 | -65.2281 | 12.9 | 13 | -89.9661 | 915.25 |
| 635 | -65.2281 | 14 | 7 | -89.9665 | 935.25 |
| 636 | -65.2281 | 14.4 | 34.7 | -89.9657 | 955.25 |
| 637 | -65.2281 | 6.6 | 15.4 | -89.9659 | 975.25 |
| 638 | -65.2281 | 9.1 | 64.7 | -89.9671 | 995.25 |
| 639 | -62.029 | 20.9 | 14.4 | -88.7649 | 875.25 |
| 640 | -62.029 | 22.9 | 7 | -88.7646 | 895.25 |
| 641 | -62.029 | 7.3 | 13.9 | -88.7656 | 915.25 |
| 642 | -62.029 | 7.5 | 14.1 | -88.766 | 935.25 |
| 643 | -62.029 | 6.3 | 37.4 | -88.765 | 955.25 |
| 644 | -62.029 | 17.3 | 46.7 | -88.7657 | 975.25 |
| 645 | -62.029 | 9.6 | 16.8 | -88.7653 | 995.25 |